\documentclass[aps,prc,twocolumn,groupedaddress,showpacs,byrevtex,amsmath,amssymb,nofootinbib]{revtex4}

\usepackage{graphicx}% Include figure files
\usepackage{dcolumn}% Align table columns on decimal point
\usepackage{bm}% bold math
\usepackage{hyperref}% add hypertext capabilities
\usepackage{comment}
\usepackage[normalem]{ulem}

\def\nuc#1#2{\relax\ifmmode{}^{#1}{\protect\text{#2}}\else${}^{#1}$#2\fi}

\newcommand{\phic}{$\phi_{\bk_I; I \mu; s \sigma}^{(+)}(\br,\xi_v,\xi_c)$}

\newcommand{\bi}{\begin{itemize}}
\newcommand{\ei}{\end{itemize}}
\newcommand{\be}{\begin{equation}}
\newcommand{\ee}{\end{equation}}

\newcommand{\bk}{{\mathbf k}}
\newcommand{\bK}{{\mathbf K}}
\newcommand{\bR}{{\mathbf R}}
\newcommand{\br}{{\mathbf r}}

\newcommand{\dif}{{\mathrm{d}}}
\newcommand{\bwt}{\begin{widetext}}
\newcommand{\ewt}{\end{widetext}}

\begin{document}

\title{Extracting three-body breakup observables from CDCC calculations with core excitations} 


%\author{}

\author{R.~de Diego}
\email{raul.martinez@tecnico.ulisboa.pt}
\affiliation{Centro de Ci\^encias e Tecnologias Nucleares, Universidade de Lisboa,
Estrada Nacional 10, 2695--066 Bobadela, Portugal}

\author{R.~Crespo}
\email{raquel.crespo@tecnico.ulisboa.pt}
\affiliation{Departamento de F\'isica, Instituto Superior T\'ecnico,
Universidade de Lisboa, Av. Rovisco Pais 1, 1049--001 Lisboa, Portugal\\ and
Centro de Ci\^encias e Tecnologias Nucleares, Universidade de Lisboa,
Estrada Nacional 10, 2695--066 Bobadela, Portugal}%

\author{A.~M.\ Moro}
\email{moro@us.es}
\affiliation{Departamento de FAMN, Facultad de F\'{\i}sica, Universidad de Sevilla, Apdo.~1065, E-41080 Sevilla, Spain}

\vspace{1cm}

\date{\today}

%\thanks{}
%\affiliation{}

\begin{abstract}
\begin{description}
\item[Background] Core-excitation effects in the scattering of two-body halo nuclei have been investigated in previous works. In particular, these effects have been found to affect in a significant way the breakup cross sections of neutron-halo nuclei with a deformed core. To account for these effects, appropriate extensions of the continuum-discretized coupled-channels (CDCC) method have been recently proposed. 
\item[Purpose] We aim to extend these studies to the case of breakup reactions measured under 
 complete kinematics or semi-inclusive reactions in which only the angular or energy distribution of one of the outgoing fragments is measured. 
\item[Method] We use the standard CDCC method as well as its extended version with core excitations (XCDCC), assuming a pseudo-state basis for describing the projectile states. Two- and three-body observables
are computed by projecting the discrete two-body  breakup amplitudes, obtained within these reaction frameworks,   onto two-body scattering states with definite relative momentum of the outgoing fragments and a definite state of the core nucleus.  

% in order to compute the angular and energy differential cross sections for each of the projectile fragments.
%weak-coupling limit  and the continuum-discretized coupled-channels method for  describing both the structure of the projectile (diagonalizing the Hamiltonian in a square-integrable basis) and the dynamics of the reaction (including the core excitation through non-central potentials).

\item[Results] Our working example is the one-neutron halo $^{11}$Be.
Breakup reactions on protons and $^{64}$Zn targets are studied at
63.7 MeV/nucleon and 28.7 MeV, respectively.
These energies, for which experimental data exist, and the targets provide two different scenarios where
the angular and energy distributions of the fragments are computed. The importance of core dynamical effects is
also compared for both cases.
\item[Conclusions] The presented method provides a tool to compute double and triple differential cross sections for outgoing fragments following the breakup of a two-body projectile, and might be useful to analyze breakup reactions with other deformed weakly-bound nuclei, for which core excitations are expected to play a role. We have found that, while dynamical core excitations are important for the proton target at
intermediate energies, they are very small for the Zn target at energies around the Coulomb barrier.  
\end{description} 
\end{abstract}

%\pacs{24.10.-i, 24.10.Eq, 25.10.+s, 25.45.De, 25.60.Gc}
 % summers
%\keywords{}

\maketitle

%---------------------------------------
\section{\label{sec:intro} Introduction}
%---------------------------------------

Recent experimental activities with nuclei in the proximity of the drip-lines have increased 
the interest in this region of the nuclear landscape.
In particular, special attention has been paid to halo nuclei,
spatial extended quantum systems with one or two loosely bound valence
particles, for example $^{11}$Be or $^{11}$Li. Breakup reactions have 
shown to be a useful tool for extracting information from these exotic structures \cite{Nak12}.
Reliable and well understood few-body reaction frameworks
to describe breakup reactions of exotic nuclei, such as \emph{haloes},
are therefore needed. Particularly,
it is timely to estimate what are the relevant excitations
mechanisms for the reaction. In the case of elastic breakup, that we address here, several few-body formalisms
have been developed to extract the corresponding cross sections:
Continuum-Discretized Coupled-Channels (CDCC) method
\cite{Raw74,Aus87}, the adiabatic approximation \cite{Ban00,Tos98},
the Faddeev/AGS equations  \cite{faddeev60,Alt},
and a variety of  semiclassical approximations
\cite{Typ94,Esb96,Kid94,Typ01,Cap04,Gar06}.
The CDCC method is based upon an explicit discretization
of all channels in the continuum and requires the
solution of an extensive set of coupling equations.
It has been applied at both low and intermediate energies.

The standard formulation of these approaches 
ignore the possible excitations of the constituent fragments: 
the states of the few-body system 
are usually described by pure single-particle configurations, 
ignoring the admixtures of different core states in the
wave functions of the projectile. These admixtures
are known to be important, particularly in the case of well-deformed
nuclei, such as the $^{11}$Be halo nucleus \cite{Lay12}. In addition,
for such two-body weakly bound system, dynamic core excitation effects in breakup  have 
been recently studied with an extension of the Distorted Wave Born 
Approximation (DWBA) formalism within a
no-recoil approximation \cite{Cre11,Mor12,Mor12b,Mor12c} and
found to be important. This method is based in the Born
approximation and ignores higher order effects (such as
continuum-continuum couplings). 
A recent attempt to incorporate 
core excitation effects within a coupled-channels calculation was done in
Ref.~\cite{Die14}, using an extended version of the CDCC formalism \cite{Summers06}, 
hereafter referred to as XCDCC. The different calculations have shown
that, for light targets, dynamic core excitations
give rise to sizable changes in the magnitude of the
breakup cross sections. Additionally, these
excitations have been found to be dominant
in resonant breakup of $^{19}$C on a proton target \cite{Lay16}.
These effects have also been studied within the Faddeev/AGS approach
in the analysis of breakup and transfer reactions \cite{Cre11,Del15,Del16}.
%by analyzing different observables in terms of the
%projectile energy or the mass of the target.}{\bf AMM: I think this statement could be written later on, because the aim of this paper is, to my understanding, to present the formalism for the calculation of 3-body observables}.

Within the CDCC and XCDCC reaction formalisms, the breakup is treated as an
excitation of the projectile to the continuum so the theoretical cross sections are described in terms of
the c.m.\ scattering angle of the projectile and the relative energy of the constituents,
using two-body kinematics. Because of this,  experimental data should be transformed to the
c.m.\ frame for comparison, but this process is ambiguous in the case of inclusive data. This is a common situation, for example, in the case of reactions involving neutron-halo nuclei in which very often only the charged fragments are detected.  Furthermore, even in exclusive breakup experiments under complete kinematics, in which this transformation is feasible, the possibility of comparing the calculated and experimental cross sections for different configurations (angular and energy) of the outgoing fragments provides a much deeper insight of the underlying processes, as demonstrated by previous analyses performed by  exclusive breakup measurements with stable nuclei \cite{Aar84}. The continuous developments at the radioactive beam facilities opens the exciting possibility of extending these studies to unstable nuclei. This will require improvements and  extensions of existing formalisms to provide these observables. 

%From the theoretical point of view, this requires     
%the breakup cross sections are  naturally expressed in terms of scattering angle of the c.m.\ of the projectile fragments and their relative energy. Experimentally, 
 
%\sout{Over the last decade, several improvements at the radioactive beam facilities have allowed complete kinematics measurements and the extraction of  exclusive observables, where the final state of the heavy fragment is identified. Under this context, accurate theoretical predictions are needed and previous works have shown their relevance when studying breakup reactions. From the exclusive cross sections we can obtain the fragment distributions and compare them with the experimental results.}

In the case of the CDCC framework,  fivefold fully exclusive cross sections were already derived  in \cite{Tos01}.
In the present work we extend this framework to the XCDCC formalism
and we provide more insight into the contribution of core admixtures (CA)
and dynamic core excitation (DCE) in the collision process.
A proper inclusion of core excitation effects in the description
of kinematically fully exclusive observables in the laboratory
frame is the primary motivation of this manuscript with the aim of pinning down the
effect of this degree of freedom in angular and energy distributions. Additionaly, this allows to 
compute  breakup observables for specific states of the core nucleus, which could be of utility in experiments using gamma rays coincidences. 
The calculations are carried out within the combined XCDCC+THO framework \cite{Die14} for
breakup reactions of $^{11}$Be on protons and $^{64}$Zn targets 
 with full three-body kinematics. The formalism is applied to investigate 
the angular and energy distributions of the $^{10}$Be fragments 
resulting from the breakup, for which new measurements have been made \cite{Pie10}.

The paper is organized as follows. In Sec.~\ref{sec:tho} we briefly
discuss the structure approach to describe two-body
loosely-bound systems with core excitation. In Sec.~\ref{sec:scatwf}, the expression of the scattering 
wave functions is derived. The three-body breakup amplitudes are shown 
in Sec.~\ref{sec:bu_amp} and the related most exclusive observables are presented 
in Sec.~\ref{sec:three_obs}. In Sec.~\ref{sec:calc}, 
we show some working examples. We study $^{11}$Be+$p$ and $^{11}$Be+$^{64}$Zn reactions at low energies
and we calculate the angular and energy distributions of the $^{10}$Be fragments 
after the breakup. In Sec.~\ref{sec:summary}, the main results and 
remarks of this work are collected.

%-------------------------------------------------
\section{\label{sec:tho}  The Structure formalism}  
%-------------------------------------------------

In this manuscript, the composite projectile is assumed to be well described by
a valence nucleon coupled to a core nucleus and the projectile states are described
in the weak-coupling limit. Thus,
these states are expanded as a superposition of
products of single-particle configurations and core states. The
energies and wavefunctions of the projectile are calculated using the
pseudo-state (PS) method \cite{Mat03}, that is, diagonalizing the model Hamiltonian
in a basis of square-integrable functions.  For the relative motion
between the valence particle and the core, we use a recently proposed 
extension of the analytical
Transformed Harmonic Oscillator (THO) basis \cite{Lay12}, which incorporates
the possible excitations of the constituents of the composite 
system. This approach differs from the binning procedure \cite{Tos01},
where the continuum spectrum is represented by a set of wave packets, constructed as a superposition of scattering states calculated by direct integration of the multi-channel Schr\"odinger equation.
The main advantage of the PS method relies on the fact
that it provides a suitable representation of the continuum
spectrum with a reduced number of functions and, as we will see below, it is particularly convenient to describe narrow resonances.

We briefly review the features of the PS basis used
in this work to describe the states of the two-body composite
projectile. The full Hamiltonian, under the weak-coupling limit, would 
be written as:
\be 
H= \hat{T}_r + V_{cv}(\br,\xi) +   H_{\rm core}(\xi) ,
\label{hpc}
\ee
where $\hat{T}_r$ (kinetic energy) and
 $V_{cv}(\br,\xi)$ (effective potential) describe
the relative motion between the core and the valence
while $H_{\rm core}(\xi)$ is the intrinsic hamiltonian of the core,
whose internal degrees of freedom are described through the coordinate
$\xi$. The eigenstates of $H_{\rm core}(\xi)$ corresponding to energies $\epsilon_{I}$
(defined by the intrinsic spin of the core, $I$) will be
denoted by $\phi_{I}$ and, additional quantum numbers, required to 
fully specify the core states, will be given below. 

%In the models considered here, the valence-core interaction is written 
%as the sum of two terms, 
%\be
%V_{vc}(\vec{r},\xi)= V_{sp}(\vec{r}) + V_{coup}(\vec{r},\xi)
%\ee

%In here, the single-particle potential $V_{sp}(\vec{r})$  describes the  motion 
%of the valence particle relative to the core, in the absence of core excitation. The following terms 
%are considered:

%\begin{eqnarray}
%V_{sp}(r)&=& V_{coul}(r)+ V^{\ell}_c(r)
%      + V^{v}_{ls}(r) \vec{\ell}\cdot \vec{s}_v 
%\nonumber \\ & + & V^{c}_{ls}(r) \vec{\ell}\cdot \vec{s}_c
%      + V_{ss}(r) \vec{s}_c \cdot \vec{s}_v 
%      + V_{ll}(r) \vec{\ell} \cdot \vec{\ell'} ,
%\label{vsp}
%\end{eqnarray}
%where $V_{coul}(\vec{r})$ is the Coulomb central potential, $V^{\ell}_c(r)$ the $\ell$-dependent nuclear part,
% $V^{v}_{ls}(r)$ and $V^{c}_{ls}(r)$ refer to the spin-orbit potentials (valence and core) while $V_{ss}(r)$
%or $V_{ll}(r)$ are the spin-spin or $\ell\cdot \ell'$ potentials, respectively.

%The coupling potential, $V_{coup}(r,\xi)$,
%depends on the core degrees of freedom (denoted generically by
%$\xi$) and it is written according to the following multipolar expansion:

%\be
%V_{coup}(\vec{r},\xi)= \sum_{\lambda } V^{\lambda}_{coup}(r,\xi) 
%Y_{\lambda \mu }(\hat r)
%\label{vcoup}
%\ee

The core-valence interaction
is assumed to contain a noncentral part, responsible for the CA in the
projectile states. In general, this potential can
be expanded into multipoles:
\be
V_{cv}(\br,\xi)= \sum_{\lambda \mu}
V_{\lambda \mu}(r,\xi) Y_{\lambda \mu}(\hat{r}) .
\label{vmult_cv}
\ee

In this work the projectile is treated within  the particle-rotor model
\cite{BM} with a permanent core deformation (assumed 
to be axially symmetric). Thus, in the body-fixed frame, the  
surface radius is parameterized in terms of the deformation parameter, $\beta_2$,
 as $R(\hat{\xi})=R_0 [1 + \beta_2  \, Y_{20}(\hat{\xi})]$, with $R_0$ an average radius.
The full valence-core interaction is obtained by deforming a central potential $V^{(0)}_{cv}(r)$ 
as follows,  
\be
V_{cv}(\br,\hat{\xi})=V_{cv}^{(0)}\left(r-\delta_2
Y_{20}(\hat{\xi})\right) , 
\label{Vcv}
\ee
where $\delta_2 = \beta_2 R_0 $ is the so-called deformation
length. The transformation to the space-fixed reference frame is made through
the rotation matrices ${\cal D}^{\lambda}_{\mu 0}(\alpha,\beta,\gamma)$
(depending on the Euler angles $\{\alpha,\beta,\gamma\}$). After
expanding in  spherical harmonics (see e.g.~Ref.~\cite{Tam65}),
the potential in Eq.~(\ref{Vcv}) reads
\be
V_{cv}(\br,\hat{\xi})= \sqrt{4 \pi} \sum_{\lambda \mu} {\cal
  V}^{\lambda}_{cv}(r) {\cal D}^{\lambda}_{\mu 0}(\alpha,\beta,\gamma)
Y_{\lambda \mu}(\hat{r}) ,
\label{vdef_cv}
\ee
with the radial form factors ($u=\cos \theta'$):
\be
{\cal V}^{\lambda}_{cv}(r) =  \frac{\sqrt{2\lambda+1}}{2} \int_{-1}^{1}
V_{cv}\left(r-\delta_2 Y_{20}(\theta',0)\right)  P_{\lambda}(u)~ \dif u .
\label{vlam}
\ee
In comparison with Eq.~(\ref{vmult_cv}) we have for a particle-rotor model:
\be
V_{\lambda \mu}^{rotor}(r,\hat{\xi})= \sqrt{4 \pi} {\cal
  V}^{\lambda}_{cv}(r) {\cal D}^{\lambda}_{\mu 0}(\alpha,\beta,\gamma) .
\label{vlm_rotor}
\ee
The eigenstates of the Hamiltonian (\ref{hpc}) 
will be a superposition of several valence configurations and core
states $\alpha=\{\ell,s,j,I\}$, with $\vec{\ell}$ (valence-core orbital angular momentum) and
$\vec{s}$ (spin of the valence) both coupled to $\vec{j}$ (total valence particle
angular momentum), for a given 
total angular momentum and parity of the composite projectile, $J^\pi$ , i.e.:
\be
\Psi_{\varepsilon, J, M}(\br,\xi) 
% =  \sum_{\alpha}^{n_\alpha} R_{\varepsilon,\alpha}(r) 
%\left[  {\cal Y}_{\ell s j}(\hat{r}) \otimes \phi_{I}(\vec{\xi}) \right]_{J M} 
= \sum_{\alpha}^{n_\alpha} \frac{R^J_{\varepsilon,\alpha}(r)}{r}  \Phi_{\alpha,J,M}(\hat{r},\xi_v, \xi_c) ,
\label{wfx}
\ee 
where $n_\alpha$ is the number of such channel configurations and the set of functions  
\be
\Phi_{\alpha,J,M}(\hat{r},\xi_v, \xi_c) \equiv \left[  {\cal Y}_{\ell s j}(\hat{r}) \otimes
   \phi_{I}(\xi) \right]_{J M} 
\label{so-basis}
\ee
is the so-called spin-orbit basis.
 
The functions $R^J_{\varepsilon,\alpha}(r)$ 
%can be calculated by direct integration of the Schr\"odinger equation. Alternatively, the eigenstates of the system can be 
are here obtained by diagonalizing the Hamiltonian in a basis of square-integrable states, such as the THO basis. For each channel $\alpha$, we consider a set of  $N$ functions $R^{THO}_{n,\alpha}(r)$ (with $n=1,\ldots,N$). The eigenvectors of the Hamiltonian will be of the form:
%
\begin{comment}
\begin{equation}
\langle \vec{r} \, \xi | n (ls) j I J M \rangle \equiv
\Phi^\alpha_{n,J M }(\vec{r},\vec{\xi}) 
 =  R^{THO}_{n,\alpha}(r) \left[  {\cal Y}_{\ell s j}(\hat{r}) \otimes
   \phi_{I}(\vec{\xi}) \right]_{J M} . 
\label{basis2}
\end{equation}
\end{comment}
%where $n$ is an index the labels the states of the basis for a given channel. 
%
%In this basis, the states of the system will be expressed as
\begin{equation}
\Psi^{(N)}_{i,J,M}(\br,\xi) 
 =  \sum_{\alpha}^{n_\alpha}  \sum_{n=1}^{N} c^i_{n,\alpha,J} \frac{R^{THO}_{n,\alpha}(r)}{r} \Phi_{\alpha,J,M}(\hat{r},\xi_v, \xi_c) ,
\label{eigenvector}
\end{equation}
where $i$ is an index that identifies each eigenstate and $c^i_{n,\alpha,J}$ 
are the corresponding expansion coefficients 
in the truncated basis, obtained by diagonalization of the full Hamiltonian 
(\ref{hpc}).

For numerical applications, the sum over the index of the THO basis can be actually performed to get
\be
\Psi^{(N)}_{i,J,M}(\br,\xi_v,\xi_c) = 
    \sum_{\alpha} \frac{g_{i,\alpha}^{J}(r)}{r} \Phi_{\alpha,J,M}(\hat{r},\xi_v, \xi_c)
\ee
where the radial function is: 
\be
g_{i,\alpha}^{J}(r) =   \sum_{n=1}^{N} c^i_{n,\alpha,J}  R^{THO}_{n,\alpha}(r) .
\ee

The negative eigenvalues of the Hamiltonian (\ref{hpc}) are identified
with the energies of bound states whereas the positive ones
correspond to a discrete representation of the continuum
spectrum.

%-----------------------------------------------------
\section{\label{sec:scatwf} Scattering wave functions}
%-----------------------------------------------------

For the calculation of the three-body scattering observables (see Sec.~\ref{sec:bu_amp}) we need also the {\it exact} scattering states of the valence+core system for a given asymptotic relative wave vector $\bm{k}_I$, and given spins of the core ($I$) and valence particle ($s$), as well as their respective projections ($\mu$ and $\sigma$, respectively), that will be denoted as \phic. These states can be written as a linear combination of the continuum states with good angular momentum $J,M$, that are of the form 
\be
\Psi^{(+)}_{\alpha,J,M}(k_I,\br,\xi_v,\xi_c) = 
   \sum_{\alpha'} \frac{f^{J}_{\alpha:\alpha'}(k_I,r)}{r} 
\Phi_{\alpha',J,M}(\hat{r},\xi_v, \xi_c) ,
\label{Psip} 
\ee
where the radial functions $f^{J}_{\alpha:\alpha'}(k_I,r)$ are the solution 
of the coupled differential equations,
\begin{eqnarray}
\left[ E_{\alpha'}-T_{r \ell'}-V^J_{\alpha':\alpha'} \right]
f^{J}_{\alpha:\alpha'}(k_I,r)= \nonumber \\
\sum_{\alpha'' \neq \alpha'}
V^J_{{\alpha':\alpha''}}f^{J}_{\alpha:\alpha''}(k_I,r) ,
\end{eqnarray}
where $E_{\alpha'}=E_{\alpha}-\epsilon_{I'}+\epsilon_{I}$, as a consequence of the energy conservation in the nucleon-core system when the latter is in the state $I$ or $I'$, $T_{r \ell'}$ is the relative kinetic energy operator, and $V^J_{{\alpha':\alpha''}}$ are the coupling potentials given by
\be
V^J_{{\alpha':\alpha''}}(r) = \langle \alpha' J M | V_{vc} | \alpha'' J M \rangle
\ee
with  $| \alpha' J M \rangle$ denoting the spin-basis defined in Eq.~(\ref{so-basis}). 
%and the nucleon in the channel ${\alpha'}$.

 \begin{comment}
{\it We use the usual coupled-channels scheme and calculate the 
eigenfunctions of the valence+core system for a given total angular momentum 
$J$ and an incoming channel $\alpha=\{\ell, s, j, I\}$. The breakup wave
functions for a relative momentum vector $\bm{k}_I$, $\phi_{\bk_I; I \mu; s \sigma}^{(+)}(\br,\xi_v,\xi_c)$
(with $\mu$, $\sigma$ denoting the spin projections), can be
expanded in terms of these eigenstates (see Appendix \ref{appendix_A} for details),
$\Psi^{(+)}_{\alpha,J,M}(k_I,\br,\xi_v,\xi_c)$, 
that are of the form:
\be
\Psi^{(+)}_{\alpha,J,M}(k_I,\br,\xi_v,\xi_c) = 
   \sum_{\alpha'} \frac{f^{J}_{\alpha:\alpha'}(k_I,r)}{r} 
\Phi_{\alpha',J,M}(\hat{r},\xi_v, \xi_c) ,
\label{Psip} 
\ee
where the set \{$\Phi_{\alpha',J,M}$\}
corresponds to the spin-orbit basis introduced in Eq.~(\ref{so-basis}) and
the radial functions $f^{J}_{\alpha:\alpha'}(k_I,r)$ are the solution 
of the coupled differential equations,
\be
\left[ E_{\alpha'}-T_{r \ell}-V^J_{{\alpha':\alpha'}} \right]
f^{J}_{\alpha':\alpha'}(k_I,r)
=\sum_{\alpha \ne \alpha'}
V^J_{{\alpha':\alpha}}f^{J}_{\alpha':\alpha}(k_I,r) .
\ee
Here $E_{\alpha'}=E-\epsilon_{I'}$ is the relative energy between the core 
and the nucleon in the channel ${\alpha'}$.
}%it 
\end{comment}
These radial functions behave asymptotically
as a plane wave in a given incoming channel $\alpha$ 
and outgoing waves in all channels, i.e.: 
\be
f^{J}_{\alpha:\alpha'}(k_I,r) \rightarrow  e^{i \sigma_\ell} 
   \left[  F_\ell(k_Ir) \delta_{\ell,\ell'} + T^{J}_{\alpha,\alpha'} H^{(+)}_{\ell'}(k_Ir) \right] ,
\label{fasym}
\ee
where  $\sigma_\ell$ are the Coulomb phase shifts, $F_\ell(k_Ir)$ the regular Coulomb function and $T^{J}_{\alpha,\alpha'}$ the T-matrix, that is directly related to the S-matrix according to:
\be
 S^{J}_{\alpha,\alpha'}   = \delta_{\alpha,\alpha'}    + 2 i T^{J}_{\alpha,\alpha'}  .
\ee

In terms of these good-angular momentum states, the scattering states result (see Appendix A) 
\begin{widetext}
\begin{align}
\label{phic}
\phi_{\bk_I; I \mu; s \sigma}^{(+)}(\br,\xi_v,\xi_c) 
%&=&\sum_{\ell,j,J,M}  C_{\ell,j,J,M} \Psi^{(+)}_{\alpha,J,M}(k_I,\br,\xi_v,\xi_c) \\
=&   \frac{4 \pi}{k_I r} \sum_{\ell,j,J,M} i^\ell  
         Y^{*}_{\ell m}(\hat {k_I})  
        \langle \ell m s \sigma | j m_j \rangle 
%\nonumber \\ 
%& \times   
\langle j m_j I \mu | J M \rangle    \sum_{\alpha'} f^{J}_{\alpha:\alpha'}(k_I,r) \Phi_{\alpha',J,M}(\hat{r},\xi_v, \xi_c) ,
\end{align}
\end{widetext}
where $m_j=M-\mu$, and $m=m_j-\sigma$.

\begin{comment}

The final states of the core+valence system will be described by
the scattering states in the form 
$(\phi_{\bk_I; I \mu; s \sigma}^{(-)}(\br,\xi_v,\xi_c))^{*}$, that can be derived
by using the time reversal operator:
%
\begin{eqnarray}
\lefteqn{(\phi_{\bk_I; I \mu; s \sigma}^{(-)}(\br,\xi_v,\xi_c))^{*}=} \nonumber \\
&&  \frac{4 \pi}{k_I r} \sum_{\substack{\ell,j,m,m_j \\ J,M }} 
      i^\ell  Y_{\ell m}(\hat {k_I})  
        \langle \ell m s \sigma | j m_j \rangle  \langle j m_j I \mu | J M \rangle  \nonumber \\
&& \times    \sum_{\alpha'} (-1)^{\ell'+I+I'}f^{J}_{\alpha:\alpha'}(k_I,r) 
\Phi^*_{\alpha',J,M}(\hat{r},\xi_v, \xi_c) .
\end{eqnarray}

\end{comment}

%---------------------------------------------------------
\section{\label{sec:bu_amp} Breakup amplitudes}
%---------------------------------------------------------
The scattering problem can be described by means of the breakup 
transition amplitude $T_{\mu \sigma; M_0}^{I s; J_0}(\bm{k}_I,\bm{K})$  
connecting an initial state $|J_0  M_0 \rangle$ with a three-body final state comprised by 
the target (assumed to be structureless), the valence particle
and the core, whose motion is described in terms of the relative momentum,
$\bm{k}_I$, and a c.m.\ wave vector, $\bm{K}$, 
that differs from the initial momentum, $\bm{K}_0$, in $|J_0  M_0 \rangle$.

We proceed to relate $T_{\mu \sigma; M_0}^{I s; J_0}$ to 
the discrete XCDCC two-body inelastic amplitudes 
$T_{M_0,M'}^{i,J_0,J'}(\theta_i,K_i)$, obtained after solving
the coupled equations in the XCDCC method and evaluated on the
discrete values of $\bm{K}$, given by the $\{\bm{K}_i\}=\{\theta_i,K_i\}$ grid.
In order to obtain this relationship, we replace the exact
three-body wave function by its XCDCC approximation in
the exact (prior form) breakup transition amplitude.
That is, we take $\Psi_{J_0,M_0}(\bm{K}_0)\simeq\Psi_{J_0,M_0}^{XCD}(\bm{K}_0)$ and
therefore we can write:
\be
T_{\mu \sigma; M_0}^{I s; J_0}(\bm{k}_I,\bm{K})\simeq
\langle \phi_{\bm{k}_I; I \mu; s \sigma}^{(-)} 
e^{i \bK \cdot \bR} | U | \Psi_{J_0,M_0}^{XCD}(\bm{K}_0)\rangle , 
\label{T-matrix1}
\ee
with the interaction $U$ between the projectile and the target
described by a complex potential expressed as follows:
\be
U=U_{ct}(\bm{r},\bm{R},\xi)+U_{vt}(\bm{r},\bm{R}) ,
\label{proj-target}
\ee
where, in addition to the projectile coordinates $\bm{r}$ and $\xi$, we
have the relative coordinate $\bm{R}$ between the projectile center of
mass and the target. Furthermore, 
the core-target interaction ($U_{ct}$) contains a
non-central part, responsible for the dynamic core excitation
and de-excitation mechanism, while the valence particle-target interaction 
($U_{vt}$) is assumed to be central. The  scattering wave functions $\phi_{\bm{k}_I; I \mu; s \sigma}^{(-)}$ are just the time reversal of 
those defined in Eq.~(\ref{phic}), and whose explicit expression is given in Appendix \ref{appendix_A}. 

Next, assuming the validity of the completeness relation in the truncated 
basis, we get:
\begin{align}
T_{\mu \sigma; M_0}^{I s; J_0}(\bm{k}_I,\bm{K}) & \simeq  \sum_{i,J',M'} \langle \phi_{\bm{k}_I; I \mu; s \sigma}^{(-)}| \Psi^{(N)}_{i,J',M'}\rangle \nonumber \\ 
& \times \langle \Psi^{(N)}_{i,J',M'}
e^{i \bK \cdot \bR} | U | \Psi_{J_0,M_0}^{XCD}(\bm{K}_0)\rangle  \nonumber \\
&= \sum_{i,J',M'} \langle \phi_{\bm{k}_I; I \mu; s \sigma}^{(-)}| \Psi^{(N)}_{i,J',M'}\rangle T_{M_0,M'}^{i,J_0,J'}(\bm{K}) ,
\label{T-matrix2}
\end{align}
where the transition matrix elements $T_{M_0,M'}^{i,J_0,J'}(\bm{K})$ 
are to be interpolated from the discrete ones $T_{M_0,M'}^{i,J_0,J'}(\theta_i,K_i)$.  
The interpolation method follows closely
the procedure of Ref.~\cite{Tos01}, with the difference that
in this reference the continuum states of the projectile
are described through a set of single-channel bins.

The overlaps between the final scattering states and the pseudo-states are explicitly given
in the Appendix \ref{appendix_B},
so Eq.~(\ref{T-matrix2}) yields  the following transition amplitude:
\begin{align}
T_{\mu \sigma; M_0}^{I s; J_0}(\bm{k}_I,\bm{K}) & \simeq
     \frac{4 \pi}{k_I} 
     \sum_{J'} \sum_{\ell, m,j} (-i)^\ell  
Y_{\ell m}(\hat {k}_I)  
     \langle \ell m s \sigma | j m_j \rangle \nonumber \\  
& \times \langle j m_j I \mu | J' M'\rangle               
      \sum_{i}{\cal G}_{\alpha}^{i,J'}(k_I)
T_{M_0,M'}^{i,J_0,J'}(\bm{K}) ,
\label{bu_ampl}
\end{align}
where
\be
{\cal G}_{\alpha}^{i,J'}(k_I)=
%\sum_{\alpha'} (-1)^{\ell+\ell'+I+I'}
\sum_{\alpha'}  \int  f^{J}_{\alpha:\alpha'}(k_I,r) g_{i,\alpha'}^{J}(r)  dr
\ee
are the overlaps between the radial parts of
the scattering states and pseudo-states wave functions. Notice  that
these overlaps are not analytical and they must be calculated
at the energies given by the relative momentum $k_I$. In practice, we compute the term involving the summation over $i$ in the r.h.s.\
of Eq.~(\ref{bu_ampl}) on a uniform momentum mesh, and interpolate this sum at the required $k_I$ values when combining them with the scattering amplitudes.
In fact,  Eq.~(\ref{T-matrix2})
is formally equivalent to the relation appearing in Ref.~\cite{Tos01} and
the main difference concerns the calculation of the overlaps. Moreover, the above
expressions can be used within the standard CDCC method (i.e. without core excitations), in which case the core internal degrees of freedom
($\xi$) are omitted.

%--------------------------------------------------------------
\section{\label{sec:three_obs} Two- and Three-body observables}
%--------------------------------------------------------------

The transition amplitudes in Eq.~(\ref{bu_ampl}),
$T_{\mu \sigma; M_0}^{I s; J_0}(\bm{k}_I,\bm{K})$
(with the relative momentum $\bm{k}_I$ and the c.m.\ wave vector $\bm{K}$),
contain the dynamics of the process for the coordinates
describing the relative and center of mass motion 
of the core and the valence particle. From these amplitudes we can
derive the two-body observables for a fixed spin of the core, $I$,
the solid angles describing the orientations of $\bm{k}_I$ ($\Omega_k$) and $\bm{K}$ ($\Omega_K$),
as well as the relative energy between the valence
and the core, $E_\mathrm{rel}$.
These observables 
factorize into the transition matrix elements and a kinematical factor:
\begin{align}
\frac{d^3\sigma^{(I)}}{d\Omega_k d\Omega_K dE_\mathrm{rel}} = & 
     \frac{\mu_{cv} k_I}{(2\pi)^5 \hbar^6} \frac{K}{K_0} \frac{\mu_{pt}^2}{2J+1} \nonumber \\
    &\times \sum_{\mu, \sigma, M_0} |T_{\mu \sigma; M_0}^{I s; J_0}(\bm{k}_I,\bm{K})|^2 ,
\label{three-obsv-cm}
\end{align}
where $\mu_{cv}$ and $\mu_{pt}$ are the valence-core and projectile-target
reduced masses. The integration over the angular part of $\bm{k}_I$
can be analytically done giving rise to the following expression for the two-body
relative energy-angular cross section distributions:
\begin{align}
\label{two-obsv}
\frac{d^2\sigma^{(I)}}{d\Omega_K dE_\mathrm{rel}} = &
     \frac{1}{2 \pi^3 \hbar^6} \frac{K}{K_0} \frac{\mu_{pt}^2 \mu_{cv}}{2J+1} \frac{1}{k_I}  \nonumber \\  
& \times  \sum_{J',M',M_0} \sum_{\ell,j} 
|\sum_{i}{\cal G}_{\alpha}^{i,J'}(k_I) T_{M_0,M'}^{i,J_0,J'}(\bm{K})|^2 .
\end{align}

This expression provides angular and energy distributions as a function of the continuous relative energy $E_\mathrm{rel}$ from the discrete (pseudo-state) amplitudes. As shown in the next section, this representation is particularly useful to describe narrow resonances in the continuum even with a small number of pseudo-states. 

The three-body observables, assuming the energy of the core is measured, are
given by \cite{Tos01}:
\begin{eqnarray}
\lefteqn{\frac{d^3\sigma^{(I)}}{d\Omega_c d\Omega_v dE_c}=
     \frac{2 \pi \mu_{pt}}{\hbar^2 K_0} \frac{1}{2J+1}}
       \nonumber \\  
&& \times \sum_{\mu, \sigma, M_0} |T_{\mu \sigma; M_0}^{I s; J_0}(\bm{k}_I,\bm{K})|^2               
      \rho(\Omega_c, \Omega_v, E_c) ,
\label{three-obsv}
\end{eqnarray}
where the phase space term $\rho(\Omega_c, \Omega_v, E_c)$, i.e., the number of states per
 unit core energy interval at solid angles $\Omega_c$ and $\Omega_v$,
 takes the form \cite{Fuc82}:
\begin{eqnarray}
\lefteqn{\rho(\Omega_c, \Omega_v, E_c)=
     \frac{m_c m_v \hbar k_c \hbar k_v}{(2\pi\hbar)^6}}
       \nonumber \\  
&& \times \left[\frac{m_t}{m_v+m_t+m_v(\bm{k}_c-\bm{K}_{tot})\cdot\bm{k}_v/k_v^2}\right] .
\label{phfac}
\end{eqnarray}

Here, the particle masses are given by $m_c$ (core), $m_v$ (valence), and $m_t$ (target)
while $\hbar \bm{k}_c$ and $\hbar \bm{k}_v$ are the core and valence particle momenta
 in the final state. The total momentum of the system corresponds to $\hbar \bm{K}_{tot}$ and
the connection with the momenta in Eq.~(\ref{bu_ampl}) is made through:

\be
\bm{K}=\bm{k}_c+\bm{k}_v-\frac{m_p}{M_{tot}}\bm{K}_{tot} ; \quad
\bm{k}_I=\frac{m_c}{m_p}\bm{k}_v-\frac{m_v}{m_p}\bm{k}_c
\ee
with $m_p=m_c+m_v$ and $M_{tot}=m_c+m_v+m_t$ the total masses of the projectile and the
three-body system, respectively.

%---------------------------------------------------------------
\section{\label{sec:calc} Application to \nuc{11}{Be} reactions}
%---------------------------------------------------------------

As an illustration of the formalism we evaluate several
 angular and energy distributions after a proper
integration of the two- and three-body observables presented in the preceding
section. In particular%As an illustration of the formalism presented in the preceding sections
, we consider the scattering of the halo nucleus $^{11}$Be on
$^1$H and $^{64}$Zn targets, comparing with data when available.
 The bound and unbound states of the
$^{11}$Be nucleus are known to contain significant admixtures  
of core-excited components \cite{For99,Win01,Cap01}, and hence core
excitation effects are expected to be important.
 
As in previous works \cite{Cre11,Die14,Che16}, the  $^{11}$Be structure is
described with the Hamiltonian of Ref.~\cite{Nun96} (model Be12-b), which consists of a 
Woods-Saxon central part ($R=2.483$~fm,
$a=0.65$~fm) and a  parity-dependent strength ($V_{c}=-54.24$~MeV for
positive-parity states and $V_{c}=-49.67$~MeV for negative-parity ones). The
potential contains also a spin-orbit term, whose radial dependence is
given by $4/r$ times the derivative of the central Woods-Saxon part, and strength
$V_\mathrm{so}=8.5$~MeV. For the $^{10}{\rm Be}$ core, this model assumes a 
permanent quadrupole deformation $\beta_2$=0.67 (i.e.~$\delta_2$=1.664~fm). Only the ground state
($0^+$) and the first excited state ($2^{+}$, $E_x= 3.368$ MeV) are
included in the model space.

%----------------------------------------------------------------------------
\subsection{$^{11}$Be+ $p$ resonant breakup \label{sec:be11p}} 
%----------------------------------------------------------------------------

We first perform a proof of principle calculation and apply the method to  the  breakup
of $^{11}$Be on a proton target at 63.7 MeV/nucleon.
Previous work \cite{Mor12} showed that the main contributions
to the total energy distribution arise from the
single-particle excitation mechanism populating the $5/2^+_1$ resonance at
$E_x$=1.78 MeV \cite{Kel12} and the contribution from the excitation of the $3/2^+_1$ resonance
($E_x$=3.40 MeV, \cite{Kel12}) due to the collective excitation  of the $^{10}$Be core. 

We repeat, in here, the calculations of Ref.~\cite{Mor12} for the angular distribution using the
 XCDCC formalism for the reaction dynamics, and the pseudo-state
basis for the structure of $^{11}$Be. Continuum states up to $J=5/2$ (both parities)
were found to be enough for convergence of the
calculated observables. These states were generated diagonalizing the $^{11}$Be Hamiltonian in a THO basis with $N=12$ radial functions
and valence-core orbital angular momenta $\ell \leq 5$. For the interaction between the projectile
and the target, Eq.~(\ref{proj-target}), we used the approximate proton-neutron Gaussian interaction as in Ref.~\cite{Mor12};
the central part of the core-target potential was calculated by a folding procedure,
using the JLM nucleon-nucleon effective interaction \cite{Jeu77} and the $^{10}$Be ground-state
density from a Antisymmetrized Molecular Dynamics (AMD) calculation \cite{Tak08}.
For the range of the JLM interaction, we used the value prescribed in the original work, $t=$1.2 fm,
and the imaginary part was renormalized by a factor $N_i=$0.8, obtained from the systematic study
of Ref.~\cite{Cor97}. The XCDCC coupled equations were integrated up to 100 fm and for total
angular momenta $J_T \leq 65$.

\begin{figure}[tb]
\begin{center}
 {\centering \resizebox*{0.85\columnwidth}{!}{\includegraphics{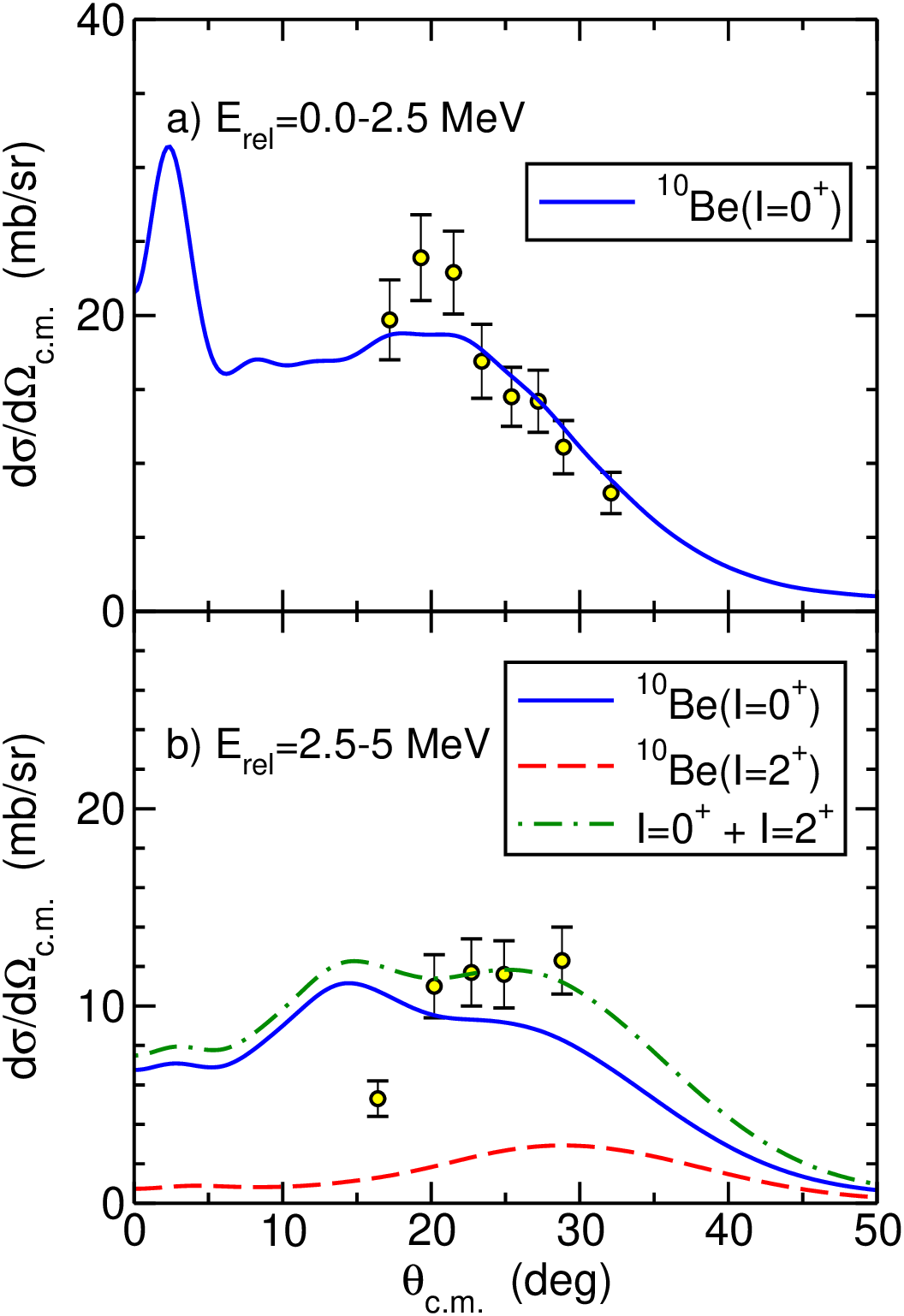}} \par}
\caption{\label{be11p_dsdw} (Color online) Differential breakup cross sections
of $^{11}$Be on protons at 63.7 MeV/nucleon, with
respect to the outgoing $^{11}$Be$^*$ c.m. scattering angle and
the neutron-core relative energy intervals $E_\mathrm{rel}=0-2.5$ MeV
(top) and $E_\mathrm{rel}=2.5-5$ MeV (bottom).
The contributions corresponding to the considered outgoing core states
are also shown.
See text for details.}
\end{center}
\end{figure}

In Fig.~\ref{be11p_dsdw} we show the two-body breakup angular distributions,
as a function of the $^{11}$Be$^*$ c.m.\ scattering angle, and we compare
the total results with the experimental data within the two
available relative energy intervals \cite{Shr04}, $E_\mathrm{rel}=0-2.5$ (top panel)
and $E_\mathrm{rel}=2.5-5$ (bottom panel). As in previous calculations
\cite{Cre11,Die14}, the agreement with
the data is fairly reasonable with the exception of
the first data point in the higher energy interval. The peak appearing at small scattering angles
for the lower energy interval is due to Coulomb breakup
and it was not present in our previous calculations due
to the smaller cutoff in the total angular momentum. We show also the separate contribution   
for each of the states of the core. 
We note that both contributions include the core excitation effect 
through the admixtures of core-excited components in the projectile
(structure effect) and the core-target potential (dynamics). However, the production of
$^{10}$Be($2^+$) is only kinematically allowed when the excitation energy is above the  
$^{10}$Be(2$^+$)+$n$ threshold, which lies at an excitation energy of 3.87~MeV with respect to the $^{11}$Be(g.s.). Consequently, 
for the lower energy interval (top panel) the system will necessarily decay into   $^{10}$Be(g.s.)+$n$, irrespective of the importance of the DCE mechanism. Notice that the emitted  $^{10}$Be($2^+$) fragments would be accompanied by the emission of a $\gamma$-ray with 
the energy corresponding to the excitation energy of this state, thus allowing an unambiguous separation of both contributions.
 
%{\bf It is worth noting that the $^{10}$Be(2$^+$) yield does not directly reflect the importance of the DCE mechanism. The reason is that, in order to produce $^{10}$Be(2$^+$) in the final state, the projectile must be excited at an excitation energy above the   $^{10}$Be(2$^+$)+$n$ threshold, which lies at an excitation energy of 3.87~MeV with respect to the $^{11}$Be(g.s.). For transitions  to continuum states below this threshold, the system will necessarily decay into   $^{10}$Be(g.s.)+$n$, irrespective of the importance of the DCE mechanism. }

\begin{figure}[tb]
\begin{center}
 {\centering \resizebox*{0.85\columnwidth}{!}{\includegraphics[angle=0]{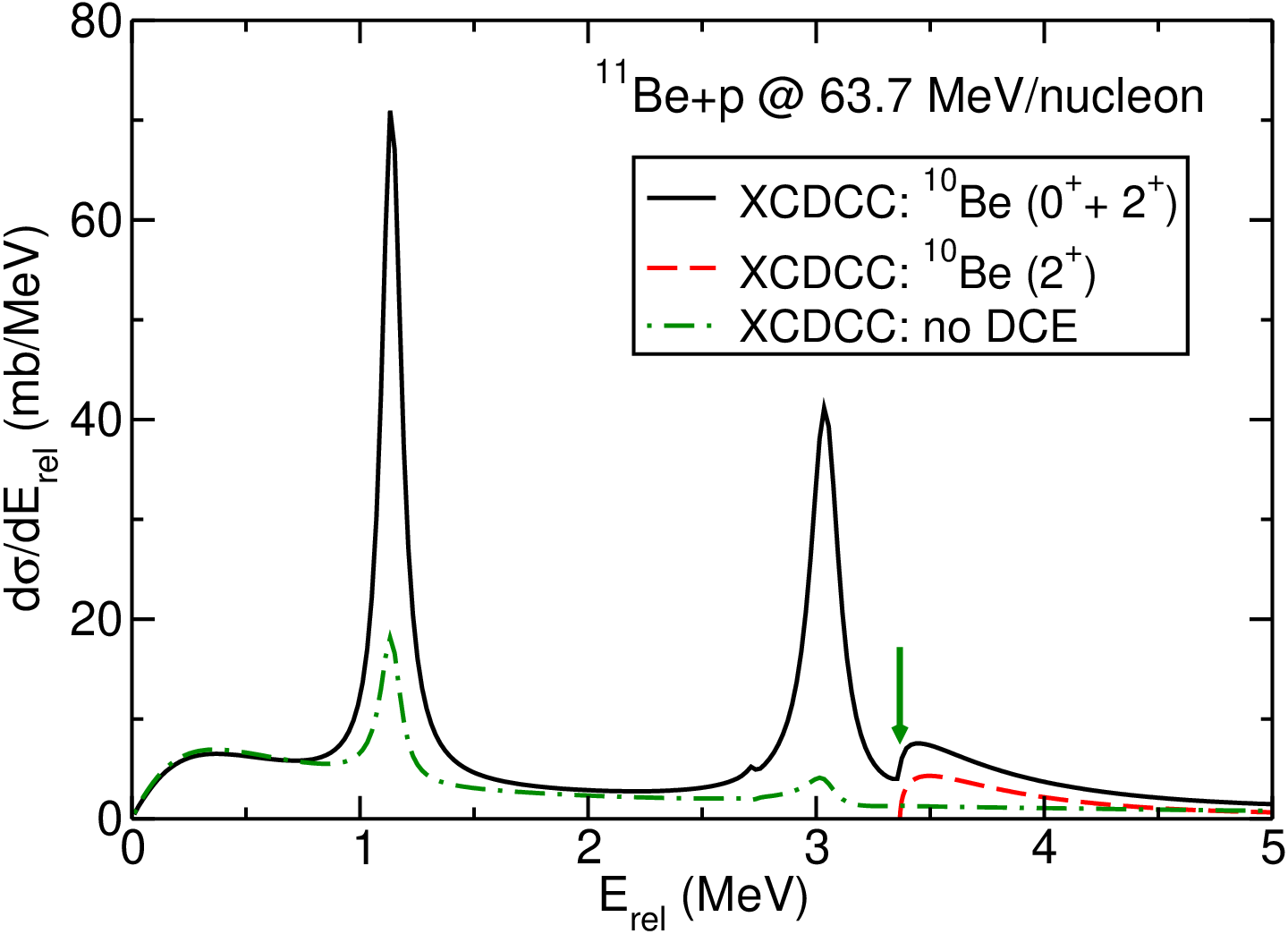}} \par}
\caption{\label{be11p_dsde} (Color online) Differential energy distribution following the breakup
of $^{11}$Be on protons at 63.7 MeV/nucleon. Solid and dashed (red)
lines correspond to the full ($0^+$+$2^+$) and the $2^+$ contribution of the XCDCC calculation while
the dot-dashed line (green) represents the result without dynamic core excitation. The arrow
indicates the energy of the $^{10}$Be(2$^+$)+$n$ threshold.}
\end{center}
\end{figure}

This is better seen in  Fig.~\ref{be11p_dsde},
where the differential energy cross section is plotted after
integration over the angular variables $\Omega_K$ in Eq.~(\ref{two-obsv}). The solid line is the full XCDCC calculation,
considering the core excitation effects in both the structure and the dynamics of the reaction, and includes the two possible final states of the $^{10}$Be nucleus. The $^{10}$Be(2$^+$) contribution (red dashed line) only appears for $E_\mathrm{rel}>3.4$~MeV, corresponding to the $^{10}$Be($2^+$)+$n$ threshold. As already noted, above this energy, the $^{10}$Be fragments can be produced in either the g.s.\ or the $2^+$ excited state. We also show
the calculation omitting the DCE mechanism (green dot-dashed curve) and considering only the CA contributions in the structure of the projectile. By comparing with the total distribution, it becomes apparent that the DCE mechanism is very important in this reaction. In particular, the energy
spectrum is dominated by two sharp peaks corresponding to the $5/2^+_1$ and $3/2^+_1$ resonances with the latter mostly populated by a DCE mechanism \cite{Cre11}. Despite the relatively small THO basis, the energy profile of these resonances is accurately reproduced and this highlights the advantage of the pseudo-state method over the binning procedure when describing narrow resonances. Finally, besides the resonant contribution, we also note that there is a non-resonant background at low relative energies and above the $^{10}$Be(2$^+$)+$n$ threshold.

%Regarding the three-body observables [Eq.~(\ref{three-obsv})], we present
%in Fig.~\ref{be11p_ds2dec} the angle-integrated
%energy distributions of the $^{10}$Be fragments in
%the laboratory frame, plotting separately the $0^+$ and $2^+$ contributions.   It is  apparent from this plot that the $^{10}$Be$(2^+)$ distribution is shifted to lower energies with respect to the $^{10}$Be(g.s.) due to the higher excitation energy required to produce the  $^{10}$Be$(2^+)$ fragments. 

\begin{figure}[tb]
\begin{center}
 {\centering \resizebox*{0.85\columnwidth}{!}{\includegraphics[angle=0]{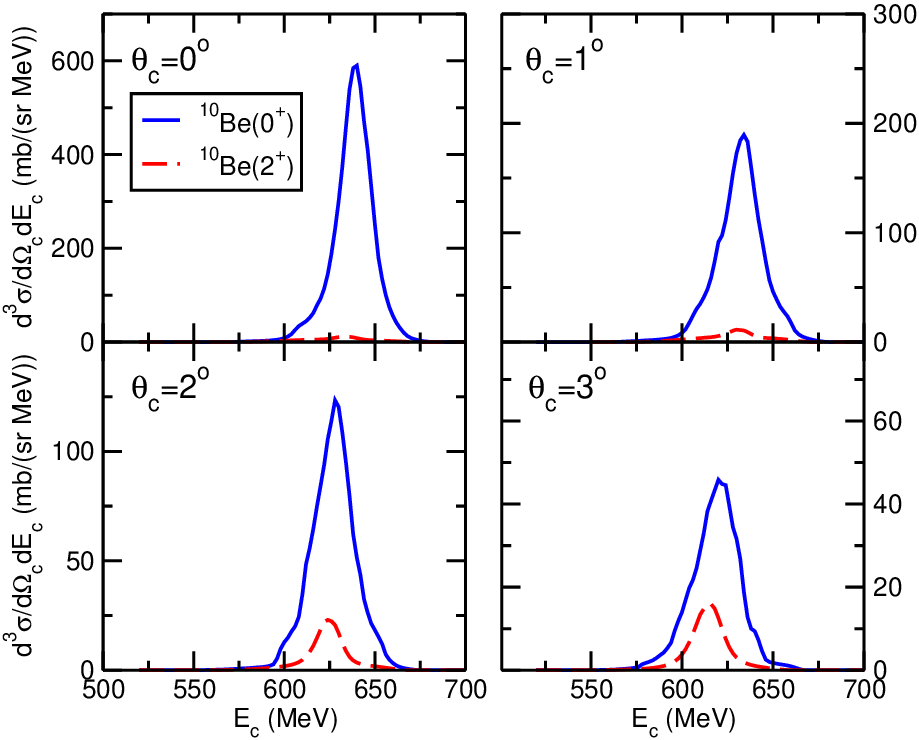}} \par}
\caption{\label{be11p_ds2dec} (Color online) Calculated laboratory-frame double differential cross section
for the $^{10}$Be fragments emitted in the process $^{11}$Be+$p$ at 63.7 MeV/nucleon
when four different scattering angles are considered.
The blue solid and red dashed  lines refer to the contributions from the different core states.}
\end{center}
\end{figure}

\begin{figure}[tb]
\begin{center}
 {\centering \resizebox*{0.85\columnwidth}{!}{\includegraphics[angle=0]{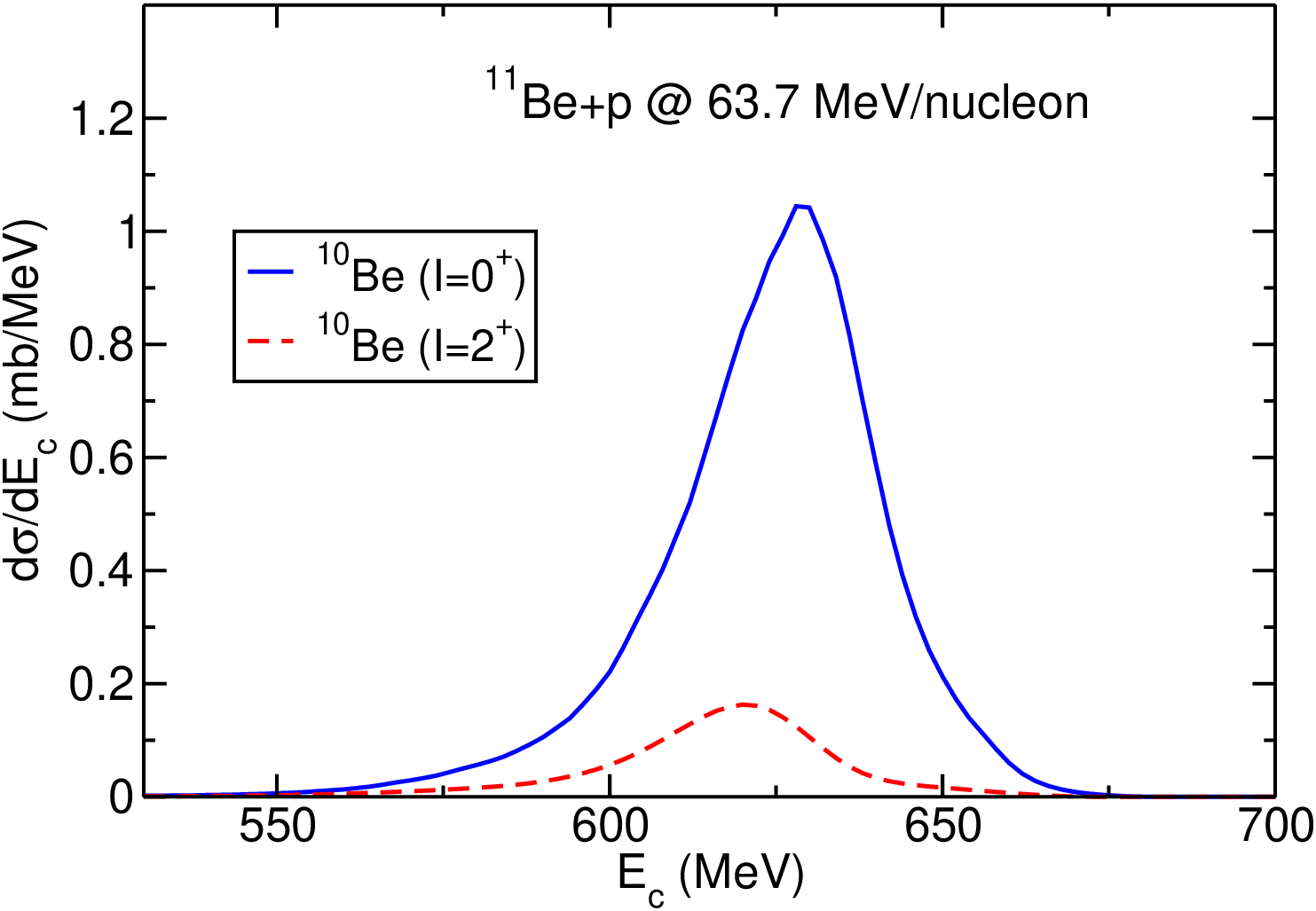}} \par}
\caption{\label{be11p_dsdec} (Color online) Calculated
differential energy cross section, as a function of the $^{10}$Be energy in the laboratory frame, for the reaction $^{11}$Be+$p$ at 63.7 MeV/nucleon. The solid (blue) and dashed (red) lines describe those calculations when the $I=0$ or $I=2$ state of the core is adopted.}
\end{center}
\end{figure}

Regarding the three-body observables [Eq.~(\ref{three-obsv})], we present
in Fig.~\ref{be11p_ds2dec} the energy distributions of the $^{10}$Be
fragments from the breakup process at four laboratory angles,
plotting separately the $0^+$ and $2^+$ contributions.
We observe from this plot the increasing relative importance of the $^{10}$Be$(2^+)$
distribution with the angle. This is expected since larger scattering angles
of the core implies a stronger interaction with the proton target. It is also apparent that this distribution
is shifted to lower energies with respect to the $^{10}$Be(g.s.)
due to the higher excitation energy required to produce the  $^{10}$Be$(2^+)$ fragments. 
The angle-integrated contributions can be seen in Fig.~\ref{be11p_dsdec}, where we 
note the dominance from the $0^+$ component to the overall energy distribution
although the $2^+$ contribution amounts to 13\% of the total cross section
at this energy.
% two contributions could be distinguished experimentally experimentally the 

%\sout{The inclusive (in neutron variables) energy distributions show a dominant contribution from the $0^+$ component although the $2^+$ cross section term is not negligible. This reveals the importance of core excitation in the breakup of $^{11}$Be on a proton target.}

%--------------------------------------------------------------------------
\subsection{$^{11}$Be+ $^{64}$Zn breakup \label{sec:be11zn}} 
%--------------------------------------------------------------------------

We consider the $^{11}$Be+$^{64}$Zn reaction at 28.7 MeV for which
inclusive breakup data have been reported in Ref.~\cite{Pie10} and
have been analyzed within the standard CDCC framework in several works
\cite{Kee10,Pie12,Dru12} and also within the XCDCC framework \cite{Die14}. The results presented here follow closely those included in this latter reference, but with two main differences: firstly, in that work the $^{10}$Be scattering angle was approximated by the $^{11}$Be$^*$ angle, assuming two-body kinematics, whereas the appropriate kinematical transformation is applied here; secondly, the XCDCC calculations are performed here in an augmented model space, including higher values of the relative orbital angular momentum between the valence and core particles. In addition, the former analysis is extended by studying the individual contributions of the $^{10}$Be $0^+$ and $2^+$ states when computing the two- and three-body observables. For the sake of comparison, we also perform standard CDCC calculations similar to those presented in \cite{Pie12} but using a larger model space, as detailed below.

For the CDCC calculations, $^{11}$Be continuum states up to $J=9/2$ (both
parities) and $J=11/2^-$ were included for maximum $n$-$^{10}$Be relative energies and orbital angular momenta $E_{rel}^\mathrm{max}=12$ MeV and $\ell_\mathrm{max}=5$, respectively. A THO basis with $N=30$ was employed and the involved interactions (i.e., the $n$-$^{10}$Be, $^{10}$Be-$^{64}$Zn, and $n$-$^{64}$Zn potentials)  were the same as in Ref.~\cite{Pie12} except for that between the neutron and $^{10}$Be. As in \cite{Pie12}, we use for this potential that from Ref.~\cite{Cap04}, but we slightly modify the depth for $\ell=2$ in order to reproduce the energy of the $5/2_1^+$ resonance obtained with the deformed model Be12-b. As for the XCDCC calculations, the following continuum states were considered: for $J \leq 5/2$ (both parities), $\ell_\mathrm{max}=3$ and  $E_{rel}^\mathrm{max}=12$ MeV. For $5/2 < J \leq 11/2$, we used $\ell_\mathrm{max}=5$ and $E_{rel}^\mathrm{max}=9$ MeV. A THO basis with $N=20$ functions was used for all $J^\pi$ with the same potentials between the projectile constituents and the target as those used in Ref.~\cite{Die14}. The coupled equations were solved in this case with the parallelized version of the {\tt Fresco} coupled-channels code \cite{Thom88}. The inclusion of high-lying excited states produces numerical instabilities in the solution of the coupled equations. A stabilization procedure similar to that proposed in \cite{Bay82} was used to get stable results.

\begin{comment}
{\bf [AMM: MAYBE WE CAN JUST REFER TO REF [DE DIEGO] FOR THE DETAILS ON THE POTENTIALS]}
For the neutron-target interaction, we adopted the optical potential used in
Ref.~\cite{Pie12}. For the $^{10}$Be+ $^{64}$Zn interaction, we
started from the optical potential derived in Ref.~\cite{Pie10}  from
a fit of the elastic scattering data for this system. To account for
the core excitation mechanism, this potential is deformed with the
same deformation length used in the structure model, i.e.,
$\delta_2=1.664$~fm. To recover the description of the $^{10}$Be+ $^{64}$Zn elastic data,
once these additional couplings are included, the optical potential
depths were readjusted, giving rise to the modified values
$V_0=-84.5$~MeV and $W_v=-34.1$~MeV, for the real and imaginary parts,
respectively. 
\end{comment} 
%The latter was originally obtained in Ref.~\cite{Pie12}, but was slightly adjusted to recover the description of the $^{10}$Be+ $^{64}$Zn elastic data, once it was deformed. 

%For the $^{11}$Be projectile, continuum states up to $J_p=9/2$ (both
%parities) and $J_p=11/2^-$ were included for energies below 12 MeV. These states were obtained by diagonalizing
%the $^{11}$Be Hamiltonian in a THO basis with $N=30$ ($N=20$)  radial functions,
%$\ell \leq 5$ ($\ell \leq 3$ when $J_p\le5/2$) and without (with) core excited states.

\begin{figure}[tb]
\begin{center}
 {\centering \resizebox*{0.85\columnwidth}{!}{\includegraphics[angle=0]{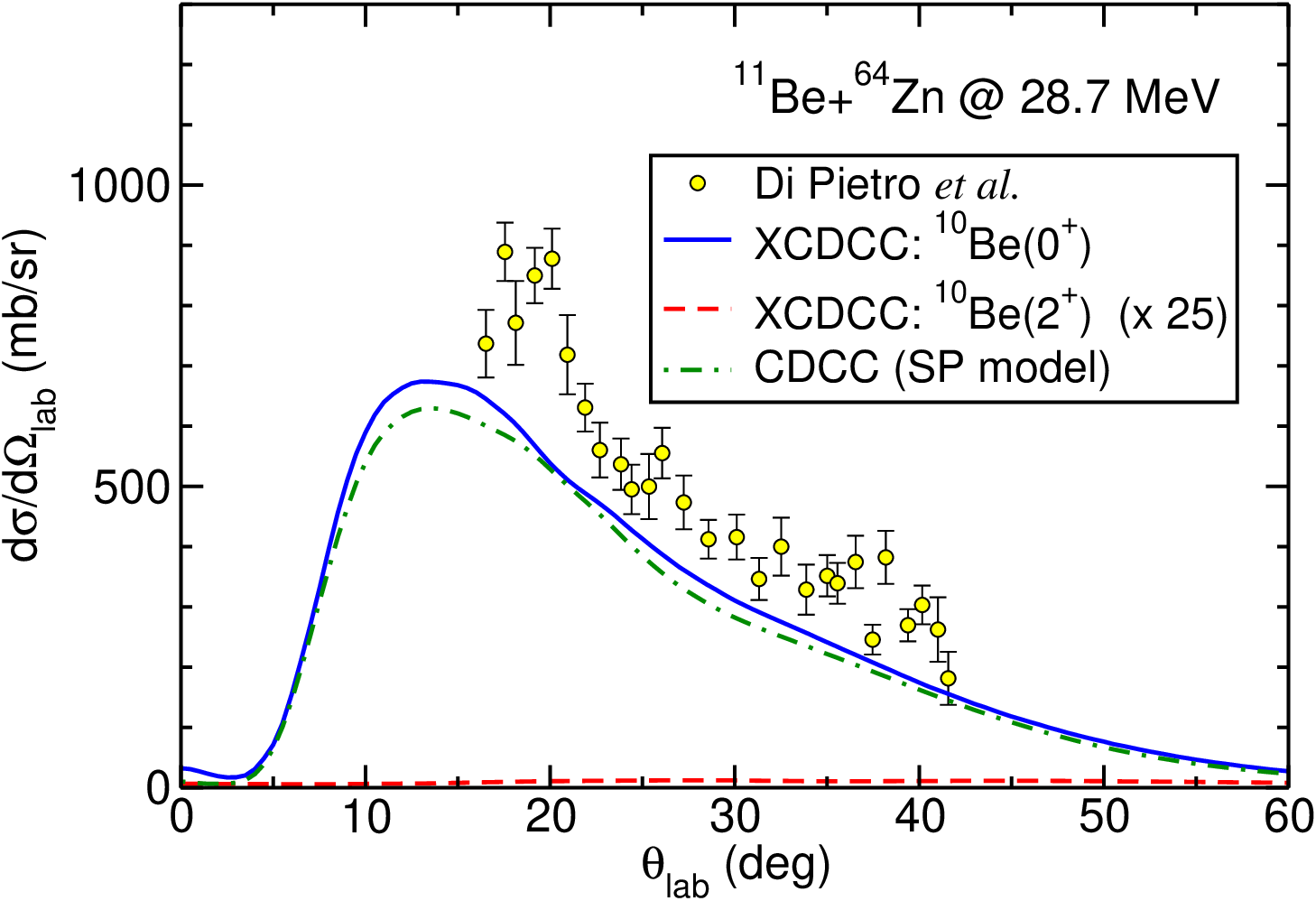}} \par}
\caption{\label{be11zn_dsdwc} (Color online) Differential cross section, as a function
of the laboratory angle, for the $^{10}$Be fragments resulting from the
breakup of $^{11}$Be on $^{64}$Zn at $E_{lab}=28.7$ MeV.
The solid line (blue) is the calculation when the core ground state, $I=0$,
is selected in Eq.~(\ref{three-obsv}) after integration
over $\Omega_v$ and $E_c$. The dashed line (red) refers to the $2^+$ contribution.
A single-particle calculation, omitting the core deformation, is also shown
as a dot-dashed curve (green). 
Experimental data are from Ref.~\cite{Pie10}.}
\end{center}
\end{figure}

In Fig.~\ref{be11zn_dsdwc} we compare the data from Ref.~\cite{Pie10} with
the present calculations. For the XCDCC calculations, the $0^+$ and $2^+$ contributions are shown separately as solid (blue) or dashed (red) lines, although the latter is found to be negligible in this case. The new CDCC calculation (green dot-dashed curve) appears to be closer to the results with core deformation but both of them clearly underestimate the experimental breakup cross section, in accordance with previous results \cite{Die14}. Nevertheless, these former calculations were carried out within a reduced model space and the proper kinematical transformation was not applied. 
%found to be larger than the CDCC result of Ref.~\cite{Pie12}, turning out closer to the
%experimental data, although a clear underestimation is still observed. 
The remaining discrepancy with the data could
be due to the contribution of non-elastic breakup events (neutron absorption or target excitation) in the data since the neutrons were not 
detected in the experiment of Ref.~\cite{Pie10}. In this regard,
it is worth recalling that the CDCC method provides only the so-called 
elastic breakup component, so the target is left in the ground state. The inclusion of these non-elastic breakup contributions has been recently the subject of several works \cite{Car15,Jin15,Pot15} but the consideration of this effect is beyond the scope of the present work. %The fact that the $2^+$ contribution is almost negligible indicates that most of the breakup is concentrated at excitation energies below the $^{10}$Be($2^+$)+$n$ threshold, as confirmed below.

%Clearly, the main contribution to the cross section comes from the $I=0$ term while
%the 2$^+$ outgoing states provides a negligible breakup angular distribution.  

\begin{figure}[tb]
\begin{center}
 {\centering \resizebox*{0.85\columnwidth}{!}{\includegraphics[angle=0]{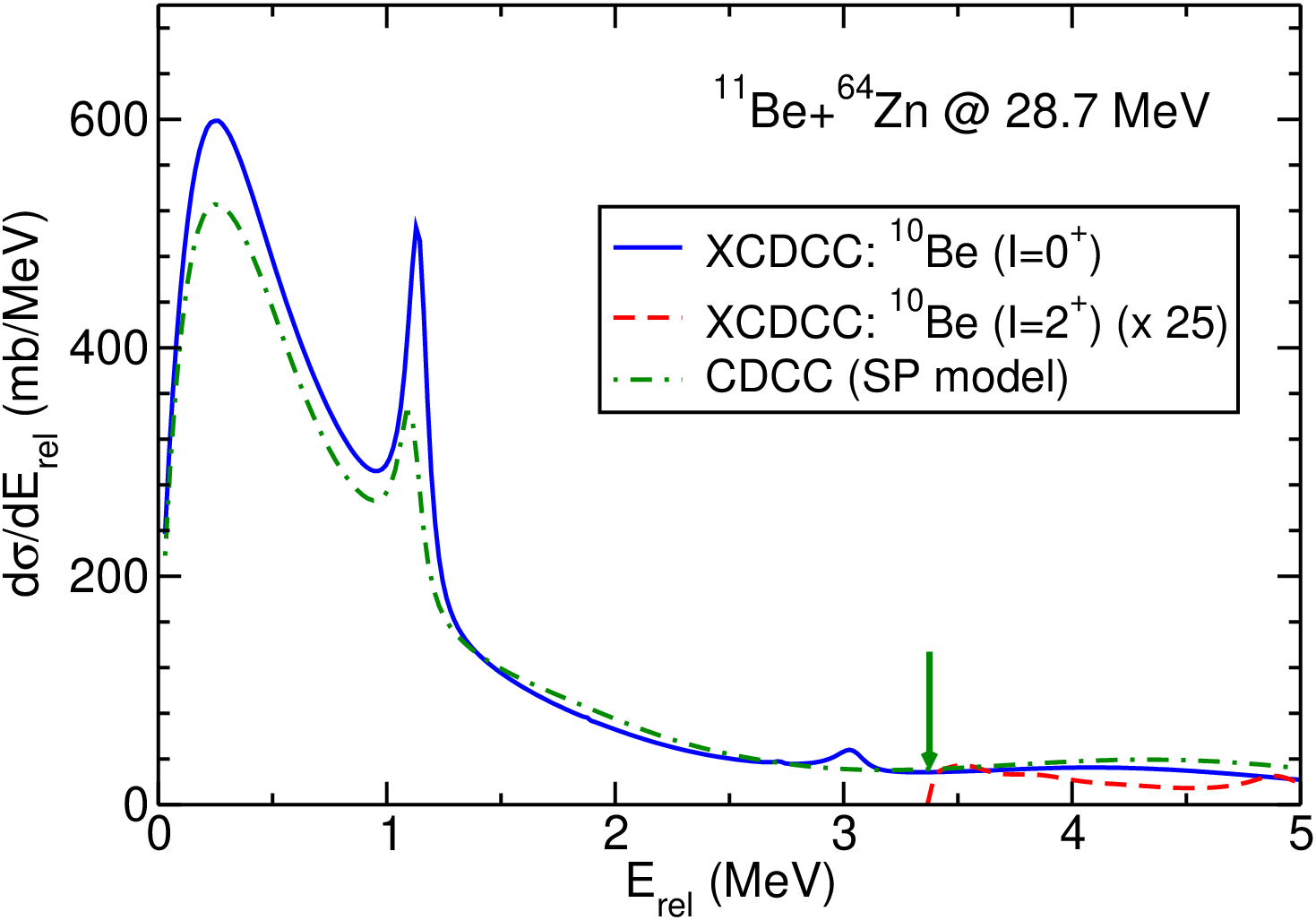}} \par}
\caption{\label{be11zn_dsde} (Color online) Calculated $^{10}$Be
differential energy cross section for the reaction $^{11}$Be+$^{64}$Zn at 28.7 MeV.
The solid (blue)  and dashed (red)  lines represent the contributions with $I=0$ or $I=2$
as the core states in the two-body distributions (\ref{two-obsv}) after
integration over $\Omega_K$. The dot-dashed (green) line considers the same calculation
without core deformation, assuming a single-particle model for the projectile
within the standard CDCC framework. The arrow indicates the energy of the $^{10}$Be(2$^+$)+$n$ threshold.}
\end{center}
\end{figure}

In Fig.~\ref{be11zn_dsde} we show the breakup cross section as a function of the $^{11}$Be excitation energy, with respect to the $^{10}$Be(g.s.)+$n$ threshold. The solid (blue) and dashed (red) lines correspond to the $^{10}$Be(g.s) and $^{10}$Be($2^+$) contributions. Most of the cross section is concentrated at low excitation energies, close to the breakup threshold, being negligible for excitation energies above the  $^{10}$Be($2^+$)+$n$ threshold. Consequently, in this reaction the $^{10}$Be will be mostly produced in its ground state. 
Moreover, it is remarkable the presence of a prominent peak at energies
around the low-lying 5/2$^+$ resonance, where the dominant component
corresponds to the $^{10}$Be(0$^+$) configuration. 
A second bump, corresponding to the population of the 3/2$^+_1$ resonance, can be barely seen. This small contribution
reflects the scarce relevance of the DCE mechanism
for this medium-mass target as pointed out in Ref.~\cite{Die14} %\sout{Consequently, the main core excitation effect arises in this case from the modification of the structure of the $^{11}$Be, rather than from the reaction dynamics.}
but, contrary to the conclusions therein, the core excitation effects in the structure of the $^{11}$Be are not so large and the assumption
of a single-particle model for the projectile yields a similar breakup cross section (green dot-dashed line). In addition, unlike the case of the proton target,
a dominant non-resonant breakup is found at low relative energies.

The different role of DCE for elastic breakup in the cases of the proton and $^{64}$Zn targets can be ascribed to the dominance of the dipole Coulomb couplings in the latter case, which hinders the effect of the quadrupole couplings associated with the excitation of the $^{10}$Be core. We expect that, for heavier targets, this dominance is enhanced and therefore  the effect of the DCE mechanism will be  further reduced. 

\begin{comment}
In the case of $^{11}$Be+$p$ breakup, we found
that the transition between the initial three-body state
and the final 2$^+$ scattering waves yields non-zero
energy differential cross sections. Now we study the effect
of these terms in the $^{11}$Be+$^{64}$Zn case. For this purpose, we compute in
Fig.~\ref{be11zn_dsde} the energy distribution of $^{10}$Be fragments in
the c.m. frame, showing a negligible contribution from the 2$^+$ states.
\end{comment}

\begin{figure}[tb]
\begin{center}
 {\centering \resizebox*{0.87\columnwidth}{!}{\includegraphics[angle=0]{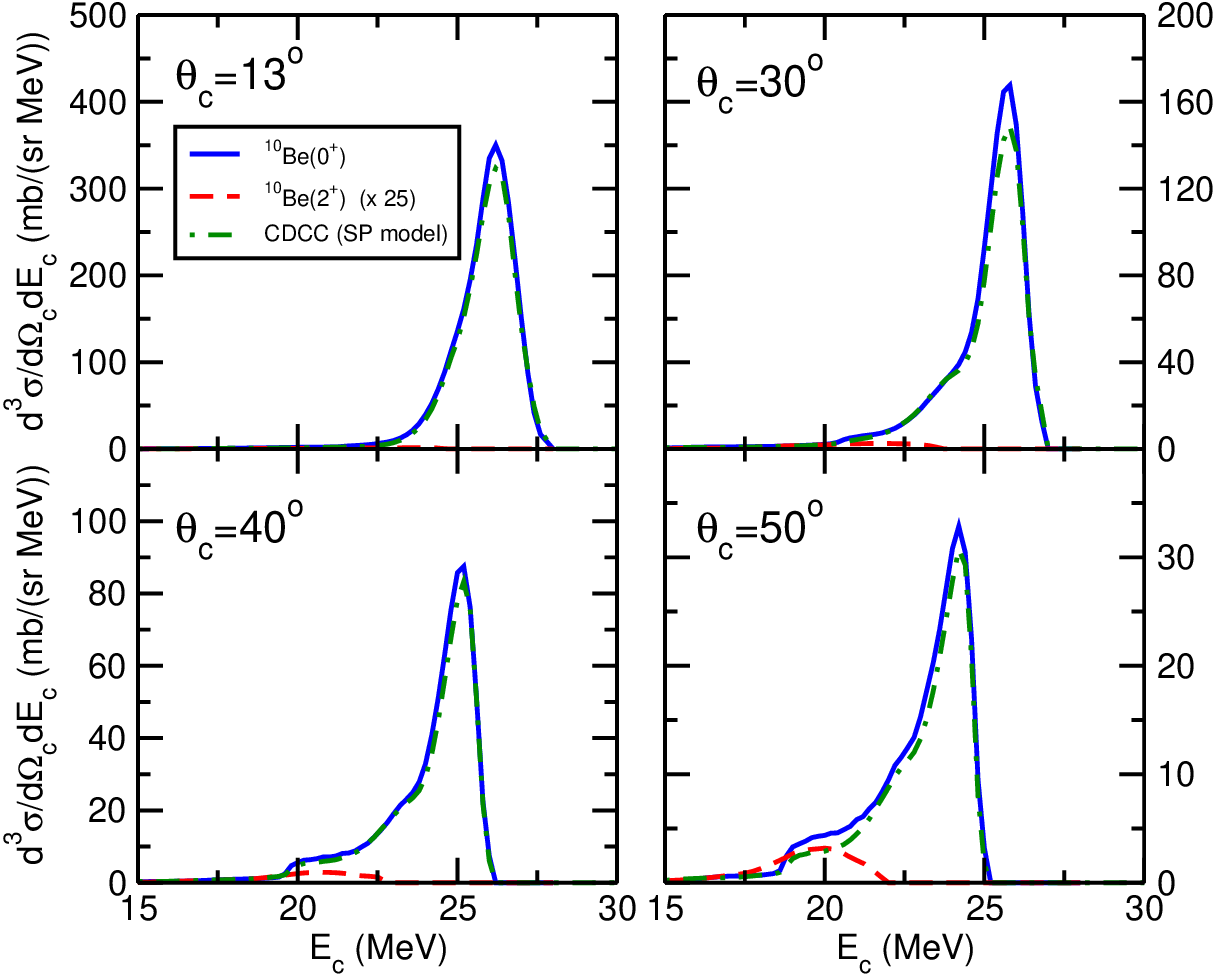}} \par}
\caption{\label{be11zn_ds2dec} (Color online) Calculated laboratory frame
$^{10}$Be cross section energy distributions for the process
$^{11}$Be+$^{64}$Zn at 28.7 MeV with different laboratory angles
in Eq.~(\ref{three-obsv}). Under the XCDCC framework we study both the
$0^+$ (blue solid) or $2^+$ (red dashed) contributions while the standard CDCC method provides
the distributions without core deformation (green dot-dashed).}
\end{center}
\end{figure}

\begin{figure}[tb]
\begin{center}
 {\centering \resizebox*{0.87\columnwidth}{!}{\includegraphics[angle=0]{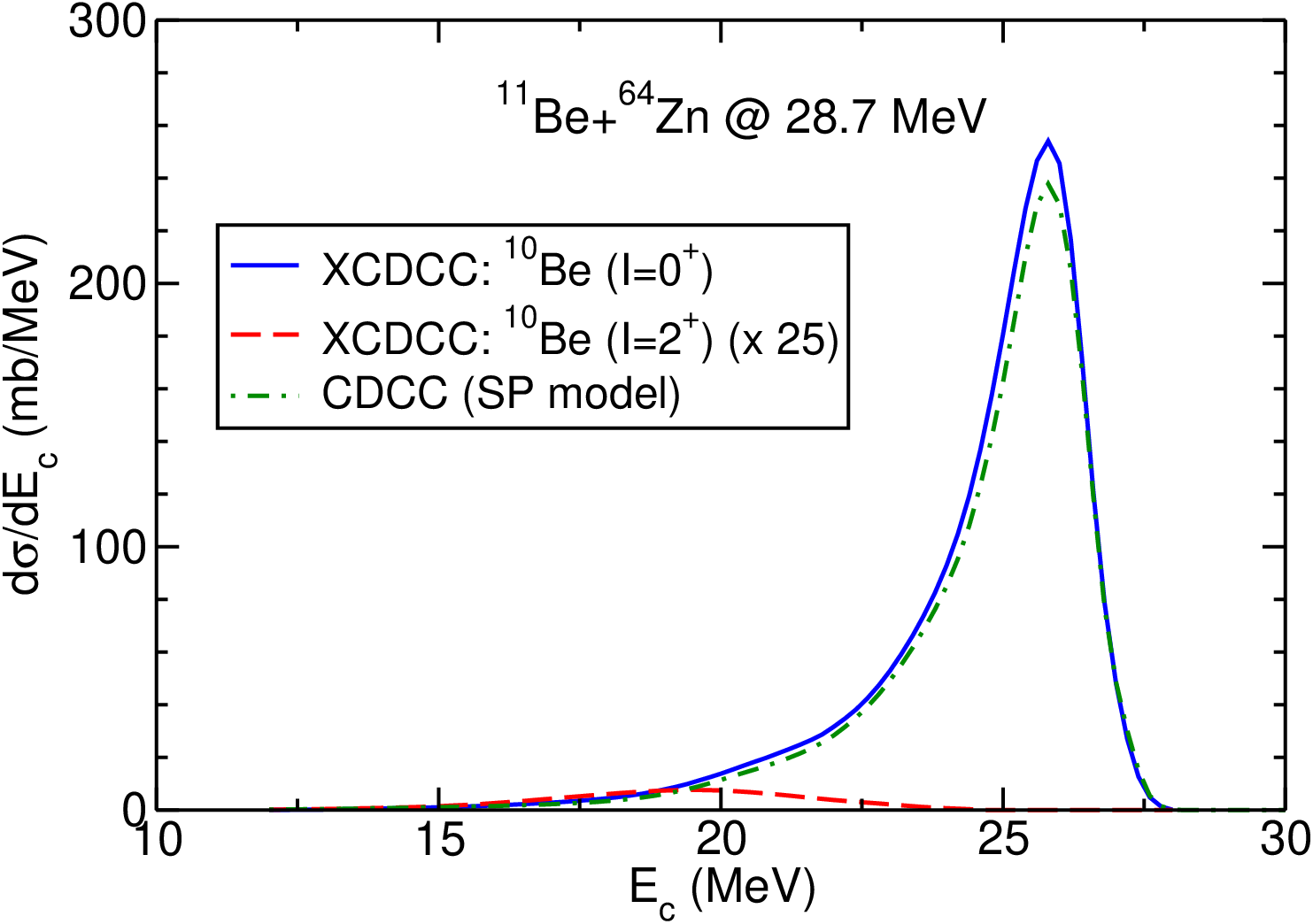}} \par}
\caption{\label{be11zn_dsdec} (Color online) Calculated laboratory frame
$^{10}$Be differential energy cross section for the breakup reaction
$^{11}$Be+$^{64}$Zn at 28.7 MeV. The blue solid and red dashed lines account for the
components obtained with the different core states in the XCDCC calculation. The distribution computed with the standard CDCC method
(no deformation) is also shown (green dot-dashed).}
\end{center}
\end{figure}

Finally, a small core excitation effect is also apparent in Fig.~\ref{be11zn_ds2dec},
where we show the breakup energy distributions of the $^{10}$Be emitted fragments
within the standard CDCC (only considering the $^{10}$Be(0$^+$) state) or XCDCC (including both 0$^+$ or 2$^+$ states in $^{10}$Be) reaction formalisms at four laboratory angles.
The relative importance of the $2^+$ state increases with increasing angle but its absolute magnitude is negligible in all cases. It is also observed an overall reduction of the mean energy of the 2$^+$ core state, as a consequence of the more negative Q-value.
The corresponding angle-integrated energy distribution is shown in Fig.~\ref{be11zn_dsdec}. As expected, this distribution is completely dominated  by the $^{10}$Be(g.s.)  contribution, displaying an asymmetric shape with a maximum around 26~MeV and a large low-energy tail extending down to 15 MeV. For core energies above the peak, the distribution exhibits a pronounced drop as a consequence of the kinematical cutoff from the energy conservation
together with the interaction between the phase
space factor and the breakup amplitude in the semi-inclusive cross section.
Actually, the sharp falloff would be present unless the breakup amplitude is very small
around the maximum energy \cite{Cre09}.
%An overall \sout{reduction} increase in the mean energy of the residual core with respect to the beam velocity is also evident as a
%consequence of the Coulomb postacceleration taking place after the dissociation of the projectile.

%--------------------------------------------------------------------
\section{\label{sec:summary} Summary and conclusions}
%--------------------------------------------------------------------

In this paper we have presented a formalism for the calculation of two- and three-body breakup observables from XCDCC calculations. The method has been applied to the case of the scattering of a two-body projectile consisting on a core and a valence particle, and taking explicitly into account core excitations. The method is based on a convolution of the discrete XCDCC scattering  amplitudes  with the exact core+valence scattering states. The formalism is a natural extension of that presented in  Ref.~\cite{Tos01} for the standard CDCC method, the main difference being the multi-channel character of the  projectile states in the present case. The convoluted transition amplitudes  are then multiplied by the corresponding phase-space factor to produce the desired (two- or three-body) differential cross sections.  The formalism provides the separate cross section for specific states of the core nucleus, thus permitting a more direct connection with experimental observables. 
 
%The formalism has been applied to the case of reactions induced by the halo nucleus $^{11}$Be, modeled as a $^{10}$Be+n two-body system, with the core either in its ground or first excited state.  
% different two- and three-body observables
%for the scattering of $^{11}$Be (composed by a core plus a 
%valence particle) on inert targets are calculated by using the combined XCDCC+THO formalism. 

%\sout{Under this methodology the effect of core excitation is present in both   the structure of the projectile, through the inclusion of core-excited components in the wave functions, and the dynamics of the reaction, by allowing core excitation and deexcitation during the collision.}  

The method has been applied to the scattering of $^{11}$Be on protons and
$^{64}$Zn. The $^{11}$Be nucleus is described in a simple  
particle-rotor model, in which the $^{10}$Be core is assumed to  have
a permanent axial deformation 
\cite{Nun96}. The core-target interaction is obtained by deforming a
central phenomenological potential.  Within the developed  approach the angular and energy distributions of
the $^{10}$Be fragments (with an intrinsic spin, $I$) following the
breakup of $^{11}$Be have been calculated and compared with experimental data, when available.

In the $^{11}$Be+$p$ reaction, we find that a significant part of the breakup cross section corresponds to the $^{10}$Be  excited state. Moreover, we have confirmed the importance of the DCE mechanism, arising
from the non-central part of the core-target interaction, for the
excitation of the low-lying $5/2^+$ and $3/2^+$ resonances \cite{Mor12}.  

We have also studied the $^{11}$Be+$^{64}$Zn reaction at 28.7 MeV, extending the previous analysis performed in Ref.~\cite{Die14}. Although in that reference  a sizable difference was observed between the calculations with and without deformation, the present calculations suggest that this difference is largely reduced if a sufficiently large model space is employed for the XCDCC calculation. 
%the effect of the core deformation was found to be very important in this reaction \cite{Die14}, the outgoing $^{10}$Be($2^+$) yield is found to be very small. This is a consequence of the fact that the breakup cross section is concentrated at relatively low excitation energies for which the decay into $^{10}$Be(2$^+$) is energetically forbidden. 
In view of these new results, we may conclude that, unlike the proton target case,  the effect of core excitation is very small in this reaction as far as the breakup cross sections concern. As a consequence, the  $^{10}$Be(2$^+$) yield is found to be negligibly small. 
 We may anticipate that this conclusion will also hold for other medium-mass or heavy systems. The qualitative difference with respect to the proton case stems from the larger importance of Coulomb couplings in the $^{64}$Zn case.

Although all the calculations presented in
this work have been performed for the $^{11}$Be nucleus, we believe
that the results are extrapolable to other weakly-bound nuclei and,
consequently, the effects discussed here should be taken in
consideration for an accurate description and interpretation of the
data. Finally, we notice that the semi-inclusive differential cross sections presented in this paper can be used to produce transverse and longitudinal momentum distributions, which have been used to obtain spectroscopic information
using both light and proton targets.
%A detailed analysis of these observables, including core excitations, will be made in a future work.}

%--------------------------------------------------------------------------
\begin{acknowledgments}
%--------------------------------------------------------------------------
 The authors would like to thank Prof. J.A.~Tostevin for useful comments and discussions.
 R.D. acknowledges support by the
 Funda\c{c}\~ao para a Ci\^encia e a Tecnologia (FCT) through Grant No.\
 SFRH/BPD/78606/2011.
 This work has also been partially supported
 by the FCT contract No. PTDC/FIS-NUC/2240/2014 as well as the Spanish
 Ministerio de  Econom\'ia y Competitividad and FEDER funds under project 
 FIS2014-53448-C2-1-P  and by the European Union's Horizon 2020 research and innovation program under grant agreement No. 654002.
\end{acknowledgments}

\appendix

%------------------------------------------------------------
\section{\label{appendix_A} Calculation of multi-channel scattering states}
%------------------------------------------------------------

Here we derive the coefficients 
$C_{\ell,j,J,M}$, that relate the scattering states with the solution of the Schr\"odinder equation for good values of $J,M$ according to the expansion
\begin{align}
\phi_{\bk_I; I \mu; s \sigma}^{(+)}(\br,\xi_v,\xi_c) = \sum_{\ell, j ,J,M}   C_{\ell,j,J,M}    \Psi^{(+)}_{\alpha,J,M}(k_I,\br,\xi_v,\xi_c). 
\end{align}
The expansion coefficients  are determined by replacing the functions $\Psi^{(+)}_{\alpha,J,M}(k_I,\br,\xi_v,\xi_c)$ by their asymptotic behaviour, Eq.~(\ref{fasym}),  
\begin{align}
\phi_{\bk_I; I \mu; s \sigma}^{(+)}(\br,\xi_v,\xi_c) & \rightarrow 
   \sum_{\ell, j ,J,M} e^{i \sigma_\ell} C_{\ell, j, J,M}   
   \sum_{\alpha'} \left [  \frac{F_\ell(k_Ir)}{r} \delta_{\ell,\ell'} \right .
 \nonumber \\    & 
+ \left . T^{J}_{\alpha,\alpha'} \frac{H^{(+)}_{\ell'}(k_Ir)}{r} \right ]
   \Phi_{\alpha',J,M}(\hat{r},\xi_v, \xi_c) \nonumber \\ 
%=\sum_{\ell,j,J,M} e^{i \sigma_\ell} C_{\ell,j,J,M}    \frac{F_\ell(k_Ir)}{r} \nonumber \\
%&\times& \Phi_{\alpha,J,M}(\hat{r},\xi_v, \xi_c) + \sum_{\ell,j,\alpha',J,M} e^{i \sigma_\ell} C^{+}_{\ell,j,J,M}    T^{J}_{\alpha,\alpha'} \frac{H^{(+)}_{\ell'}(k_Ir)}{r} \nonumber \\
%&\times& \Phi_{\alpha',J,M}(\hat{r},\xi_v, \xi_c) ,
\label{phi_asym_coefs}
\end{align}
where, in the r.h.s. of the equation, we have separated for convenience the part containing the regular Coulomb function. 

To compare this with the asymptotic behaviour we need the partial-wave decomposition of the plane wave:
\be
e^{i \bk_I \cdot \br}= \frac{4 \pi}{k_I r} \sum_{\ell,m} 
             i^\ell (k_Ir) j_{\ell}(k_Ir)Y^{*}_{\ell m}(\hat{k_I}) Y_{\ell m}(\hat{r}) .
\ee
More generally, in presence of Coulomb, the expansion above becomes:
\be
\chi_C({\bk_I, \br})= \frac{4 \pi}{k_I r} \sum_{\ell,m} i^\ell  e^{i \sigma_\ell}
              F_{\ell}(k_Ir) Y^{*}_{\ell m}(\hat{k_I})  Y_{\ell m}(\hat{r}) .
\ee

Using this result, the plane wave reads:
\begin{eqnarray}
\lefteqn{\chi_C({\bk_I, \br})\varphi^{(v)}_{s\sigma}(\xi_v) \varphi^{(c)}_{I\mu}(\xi_c)=} \\
&&             \frac{4 \pi}{k_I r} \sum_{\ell,m}  i^\ell e^{i \sigma_\ell}
              F_{\ell}(k_Ir) Y^{*}_{\ell m}(\hat{k_I})  Y_{\ell m}(\hat{r}) 
              \varphi^{(v)}_{s\sigma}(\xi_v)\varphi^{(c)}_{I\mu}(\xi_c) . \nonumber
\end{eqnarray}
In oder to express this state in terms of the basis (\ref{so-basis}), we
 use the following expansions:
\be
Y_{\ell m}(\hat{r})  \varphi^{(v)}_{s\sigma}(\xi_v) = 
   \sum_{j,m_j} \langle \ell m s \sigma | j m_j \rangle   {\cal Y}_{(\ell s) j m_j}(\hat{r},\xi_v)
\ee
 
\be
 {\cal Y}_{(\ell s) j m_j}(\hat{r},\xi_v) \varphi^{(c)}_{I\mu}(\xi_c) =
    \sum_{J M} \langle j m_j I \mu | J M \rangle \Phi_{\alpha,J,M}(\hat{r},\xi_v, \xi_c) .
\ee
 
So, collecting results,
\begin{widetext}
\begin{align}
\chi_C({\bk_I ,\br})\varphi^{(v)}_{s\sigma}(\xi_v) \varphi^{(c)}_{I\mu}(\xi_c) & =
\frac{4 \pi}{k_Ir} \sum_{\ell,m} Y^{*}_{\ell m}(\hat {k_I}) i^\ell e^{i \sigma_\ell} 
  F_\ell(k_Ir)  \sum_{j,m_j} \sum_{J,M} \langle \ell m s \sigma | j m_j \rangle 
        \langle j m_j I \mu | J M \rangle \Phi_{\alpha,J,M}(\hat{r},\xi_v, \xi_c) .
\end{align}
\end{widetext}
The above expression gives the plane-wave part of Eq.~(\ref{phi_asym_coefs}), so we get for the $C$ coefficients:
\be
C_{\ell, j,J,M}= \frac{4 \pi}{k_I} \sum_{m, m_j} i^\ell Y^{*}_{\ell m}(\hat {k_I})  
        \langle \ell m s \sigma | j m_j \rangle  \langle j m_j I \mu | J M \rangle .
\label{C_coefs}
\ee

Therefore, the scattering states $\phi_{\bk_I; I \mu; s \sigma}^{(+)}(\br,\xi_v,\xi_c)$ are expressed as follows:
\begin{eqnarray}
\label{phi_coefs}
\lefteqn{\phi_{\bk_I; I \mu; s \sigma}^{(+)}(\br,\xi_v,\xi_c)} \nonumber \\
%&=&\sum_{\ell,j,J,M}  C_{\ell,j,J,M} \Psi^{(+)}_{\alpha,J,M}(k_I,\br,\xi_v,\xi_c) \\
&=&   \frac{4 \pi}{k_I r} \sum_{\ell,j,J,M} i^\ell  
         Y^{*}_{\ell m}(\hat {k_I})  
        \langle \ell m s \sigma | j m_j \rangle  \langle j m_j I \mu | J M \rangle \nonumber \\ 
&&\times     \sum_{\alpha'} f^{J}_{\alpha:\alpha'}(k_I,r) \Phi_{\alpha',J,M}(\hat{r},\xi_v, \xi_c) ,
\end{eqnarray}
where $m_j=M-\mu$, and $m=m_j-\sigma$.

The scattering states appearing in the T-matrix amplitude, Eq.~(\ref{T-matrix1}), are $(\phi_{\bk_I; I \mu; s \sigma}^{(-)}(\br,\xi_v,\xi_c))^{*}$, which can 
be derived from $\phi_{\bk_I; I \mu; s \sigma}^{(+)}(\br,\xi_v,\xi_c)$ by application of the time-reversal operator,
% $(\phi_{\bk_I; I \mu; s \sigma}^{(-)}(\br,\xi_v,\xi_c))^{*} = {\bf PHASES (-1) ^{I-\mu}} \phi_{-\bk_I; I -\mu; s \sigma}^{(+)}(\br,\xi_v,\xi_c)$, 
resulting \cite{How08}, 
%final states of the core+valence system will be described by
%the scattering states in the fo
%$(\phi_{\bk_I; I \mu; s \sigma}^{(-)}(\br,\xi_v,\xi_c))^{*}$, that can be derived
%by using the time reversal operator:
%
\begin{widetext}
\begin{align}
\label{eq:phimc}
(\phi_{\bk_I; I \mu; s \sigma}^{(-)}(\br,\xi_v,\xi_c))^{*}=    
% \nonumber \\ && 
 \frac{4 \pi}{k_I r} \sum_{\substack{\ell,j,m,m_j \\ J,M }} 
      i^\ell  Y_{\ell m}(\hat {k_I})  
        \langle \ell m s \sigma | j m_j \rangle  \langle j m_j I \mu | J M \rangle 
% \nonumber \\ && \times  
  \sum_{\alpha'} (-1)^{\ell'+I+I'}f^{J}_{\alpha:\alpha'}(k_I,r) 
\Phi^*_{\alpha',J,M}(\hat{r},\xi_v, \xi_c) .
\end{align}
\end{widetext}

%-----------------------------------------------
\section{\label{appendix_B} Overlap functions }
%-----------------------------------------------
Starting from the states ($\ref{eq:phimc}$), and writing the THO eigenstates for a given angular momentum and projection $J',M'$ as:
\be
\Psi^{(N)}_{i,J',M'}(\br,\xi_v,\xi_c) = 
    \sum_{\alpha''} \frac{g_{i,\alpha''}^{J'}(r)}{r} \Phi_{\alpha'',J',M'}(\hat{r},\xi_v, \xi_c) ,
\ee

the overlap between them yields
%Using the explicit expression of the $C$ coefficients (\ref{C_coefs}), the scattering overlaps are written as
\begin{align}
\langle \phi_{\bm{k}_I; I \mu; s \sigma}^{(-)} | \Psi^{(N)}_{i,J',M'}\rangle
  & =      
 \frac{4 \pi}{k_I} 
      \sum_{\ell, m, j} (-i)^\ell   Y_{\ell m}(\hat {k_I})  \langle \ell m s \sigma | j m_j \rangle  \nonumber \\ 
    & \times  \langle j m_j I \mu | J' M'\rangle    
   {\cal G}_{\alpha}^{i,J'}(k_I) ,    
\end{align}
which, in addition to geometric coefficients, contain the function 
${\cal G}_{\alpha}^{i,J'}(k_I)=\sum_{\alpha'} (-1)^{\ell+\ell'+I+I'} {\cal O}_{\alpha,\alpha'}^{i,J'}(k_I)$, 
with the $ {\cal O}_{\alpha,\alpha'}^{J}(k_I)$ representing the overlaps between the radial functions:
\be
 {\cal O}_{\alpha,\alpha'}^{i,J}(k_I) = \int  f^{J}_{\alpha:\alpha'}(k_I,r) g_{i,\alpha'}^{J}(r)  dr .
\label{overlaps_tho}
\ee

We also note that the total parity of the pseudo-states, $\Psi^{(N)}_{i,J',M'}(\br,\xi_v,\xi_c)$,
is given by the factor $(-1)^{\ell+I}$ and, consequently,
the phase appearing in the function ${\cal G}_{\alpha}^{i,J'}(k_I)$ is equal to one.

\bibliography{xcdcc}

\end{document}